\documentclass[a4paper]{jpconf}
\usepackage{graphicx,iopams,amsmath,cite}
\begin{document}
\title{Equation of state for tungsten over a wide range of~densities and internal energies}

\author{K~V~Khishchenko}

\address{Joint Institute for High Temperatures of the Russian Acad\-emy of Sciences, Izhorskaya 13 Bldg 2, Moscow 125412, Russia}

\ead{konst@ihed.ras.ru}

\begin{abstract}
A caloric model, which describes the pressure--density--internal-energy relationship in a broad region of condensed-phase states, is applied for tungsten. As distinct from previously known caloric equations of state for this material, a new form of the cold-compression curve at $T = 0$~K is used. Thermodynamic characteristics along the cold curve and shock Hugoniots are calculated for the metal and compared with some theoretical results and experimental data available at high energy densities.
\end{abstract}



Tungsten and its alloys are widely used in technologies as structural materials working under extreme conditions involving high mechanical and thermal influences~\cite{Bushman-Fortov-Kanel-Ni-1993, Russkikh-Baksht-Labetskii-Levashov-Tkachenko-Khishchenko-Shishlov-Fedyunin-Chaikovskii-PPR-2006-eng, Rousskikh-Baksht-Chaikovsky-Fedunin-Khishchenko-Labetsky-Levashov-Shishlov-Tkachenko-2006, Senchenko-Belikov-Popov-2015}. An equation of state (EOS) for tungsten is of interest to use in hydrodynamic simulations of processes at high energy densities \cite{Rakhel-Korobenko-Savvatimski-Fortov-IJT-2004, Oreshkin-Baksht-Labetsky-Rousskikh-Shishlov-Levashov-Khishchenko-Glazyrin-TP-2004-eng, Tkachenko-Khishchenko-Levashov-IJT-2005, Povarnitsyn-Khishchenko-Levashov-IJImpEng-2006, Tkachenko-Levashov-Khishchenko-JPA-2006, Oreshkin-Khishchenko-Levashov-Rousskikh-Chaikovskii-HT-2012, Andreev-Povarnitsyn-Veysman-Faenov-Levashov-Khishchenko-Pikuz-Magunov-Rosmej-Blazevic-Pelka-Schaumann-Schollmeier-Roth-LPB-2015}. In this work, a caloric EOS $P = P(V, E)$, which provides an adequate description of thermodynamic properties of tungsten in condensed-phase states over a wide range of densities and internal energies, is presented.
Here, $P$ is the pressure, $V=\rho^{-1}$ is the specific volume, $\rho$ is the density, $E$ is the specific internal energy.
In contrast to the EOSs derived for this metal previously \cite{Kormer-Urlin-Popova-1961-eng, Bushman-Fortov-Lomonosov-1991-proc, Bushman-Lomonosov-Fortov-1992-metals-eng, Lomonosov-Bushman-Fortov-Khishenko-1994, Lomonosov-Thesis-2000, Tkachenko-Khishchenko-Vorobev-Levashov-Lomonosov-Fortov-HT-2001-eng, Lomonosov-Fortov-Khishchenko-Levashov-2002-article, Fortov-Lomonosov-PhysUsp-2014}, a method \cite{Khishchenko-TPL-2004-eng} for calculation of the curve of cold compression at $T = 0$~K is used.

The EOS model is formulated in the general form as
\begin{equation}\label{PVE}
P(V,E)=P_c(V)+\frac{\Gamma(V,E)}{V}\left[E-E_c(V)\right],
\end{equation}
where $E_c(V)$ and $P_c(V)=- \mathrm{d}E_c/\mathrm{d}V$ are the cold components of energy and pressure at $T=0$~K, and $\Gamma(V,E)$ is a coefficient determining the contribution of thermal components of EOS.

The cold interaction energy in the compression region ($\sigma_c \geqslant 1$, where $\sigma_{c} = V_{0c}/V$, $V_{0c}$ is the specific volume at $P=0$ and $T=0$~K) is given by the relation \cite{Khishchenko-TPL-2004-eng}
\begin{equation}\label{Ecs1} 
E_c(V) = V_{0c} a_0 \ln\sigma_c - V_{0c} \sum_{i=1}^{6} \frac{3a_i}{i} \Big( \sigma_c^{-i/3} - 1 \Big) + V_{0c} \sum_{i=1}^{2} \frac{3b_i}{i} \Big( \sigma_c^{i/3} - 1 \Big), 
\end{equation}
providing for the condition
\begin{equation}
E_c(V_{0c}) = 0.\label{E0c}
\end{equation}
As can be readily seen, differentiation of the energy (\ref{Ecs1}) with respect to volume yields an equation for the pressure $P_c(V)$, which is analogous to the relation proposed previously \cite{Kalitkin-Govorukhina-1965} as an expansion of the Thomas--Fermi model in powers of the atomic cell radius $r_c\sim\sigma_c\!^{-1/3}$.

The value of coefficient $b_2$ in equation~(\ref{Ecs1}) is determined from the condition of coincidence with the model of degenerate ideal Fermi-gas of non-relativistic electrons \cite{Landau-Lifshitz-V-1980-eng} at high compression ratios $\sigma_c \gtrsim 10^3$--$10^4$,
\begin{equation}
b_2=Z^{5/3}\frac{1}{5}\left(3\pi^2\right)^{2/3}a_B^2E_H\left(Am_uV_{0c}\right)^{-5/3},
\end{equation}
where $E_H$ is the Hartree energy,
$a_B$ is the Bohr atomic radius,
$m_u$ is the atomic mass unit (amu),
$A$ is the atomic mass (in amu),
$Z$ is the atomic number of the element.

In order to find the coefficients $b_1$ and $a_i$ in equation~(\ref{Ecs1}), one need solve the problem of minimization of the root-mean-square deviation of pressure at some values of volume in the range $\sigma_c=50$--$10^3$ from the results of calculation using the Thomas--Fermi model with quantum and exchange corrections \cite{Kalitkin-Kuzmina-1975-eng} taking into account the conditions for the pressure, bulk modulus and its derivative with respect to pressure at $\sigma_c=1$:
\begin{gather}
P_c(V_{0c})=0,\label{P0c}\\
B_c(V_{0c})=-V\mathrm{d}P_c/\mathrm{d}V=B_{0c},\label{B0c}\\
B'_c(V_{0c})=\mathrm{d}B_c/\mathrm{d}P_c=B'_{0c}\label{B1c}.
\end{gather}
The problem of conditional minimization is solved with the introduction of Lagrange factors \cite{Bushman-Fortov-Lomonosov-1991-proc}. The values of the parameters $V_{0c}$, $B_{0c}$ and $B'_{0c}$ are fitted by iterations so as to satisfy under normal conditions the value of specific volume $V_0$ and the values of isentropic compression modulus $B_S=-V(\partial P/\partial V)_S=B_{S0}$ and its pressure derivative $B'_S=\left(\partial B_S/\partial P\right)_S=B'_{S0}$ determined from the data of static and shock compressibility measurements.

The energy on the cold curve in the rarefaction region ($\sigma_c<1$) is given by the polynomial
\begin{equation}\label{Ecs2}
E_c(V)  =  V_{0c} \bigg(\sigma_c^m \frac{ a_m }{m} + \sigma_c^n \frac{a_n}{n} - \sigma_c^l \frac{a_m +a_n}{l}\bigg) + E_{sub},
\end{equation}
which provides for the value of the sublimation energy $E_c=E_{sub}$ at $V\to \infty$ as well as for the condition~(\ref{P0c}).
The requirement to satisfy equations~(\ref{E0c}), (\ref{B0c}) and (\ref{B1c}) leaves only two free parameters ($l$ and $n$) in equation~(\ref{Ecs2}).

The functional dependence of the coefficient $\Gamma$ upon the volume and internal energy is defined analogously to the caloric model~\cite{Bushman-Fortov-Lomonosov-1991-proc} in the following form:
\begin{equation}\label{Gamma}
   \Gamma(V,E)=\gamma_i+\frac{\gamma_c(V)-\gamma_i}{1+\sigma^{-2/3}\left[E-E_c(V)\right]/E_a},
\end{equation}
where $\sigma=V_{0}/V$, $\gamma_c(V)$ corresponds to the case of low thermal energies, and $\gamma_i$ characterizes the region of highly heated condensed substance. The anharmonicity energy $E_a$, which sets the thermal energy of a transition from one limiting case to another, is determined from the results of shock-wave experiments at high pressures.

The volume dependence of the cold component of $\Gamma$ is defined as \cite{Bushman-Lomonosov-Fortov-Khishchenko-Zhernokletov-Sutulov-JETP-1996}
\begin{equation}\label{gamma_c}
\gamma_c(V)=2/3+\left(\gamma_{0c}-2/3\right)\frac{\sigma_n^2+\ln^2 \sigma_m}{\sigma_n^2+\ln^2 (\sigma/\sigma_m)},
\end{equation}
where
\begin{equation}\label{gamma_0c}
\gamma_{0c}=\gamma_i+\left(\gamma_0-\gamma_i\right)\bigg[1+\frac{E_0-E_c(V_0)}{E_a}\bigg]^2, 
\end{equation}
$E_0$ and $\gamma_0$ are the values of specific internal energy and Gr\"uneisen coefficient $\gamma=V(\partial P/\partial E)_V$ under normal conditions. The form of $\gamma_c(V)$ ensures validity of the condition $\gamma(V_0,E_0)=\gamma_0$, and gives the asymptotic value $\gamma_c=2/3$ in the limiting cases of low and high compression ratios. The parameters $\sigma_n$ and $\sigma_m$ are determined from the requirement of optimum fit to experimental data on shock compressibility of porous samples of a substance in question.

The coefficients of the equation of state for tungsten within the framework of the model (\ref{PVE})--(\ref{gamma_0c}) are as follows:
$V_0 = 0.05202$,
$V_{0c} = 0.05188$,
$a_0 = 357809.521$,
$a_1 = -80307.729$,
$a_2 = -1209123.702$,
$a_3 = 2665469.845$,
$a_4 = -2635467.484$,
$a_5 = 1294811.638$,
$a_6 = -256243.391$,
$b_1 = -167421.703$,
$b_2 = 30473.005$,
$a_m = 135.91$,
$a_n = -13.913$,
$m = 4.436$,
$n = 12$,
$l = 1$,
$E_{sub} = 4.8$,
$\gamma_{0c} = 1.85$,
$\sigma_m = 0.84$,
$\sigma_n = 0.4$,
$\gamma_i = 0.5$ and 
$E_a = 15$.
The units of measure correspond to the initial units of $P = 1$~GPa, $V = 1$~cm$^3$/g and $E = 1$~kJ/g.

\begin{figure}[t]
\centering\includegraphics[width=0.75\columnwidth]{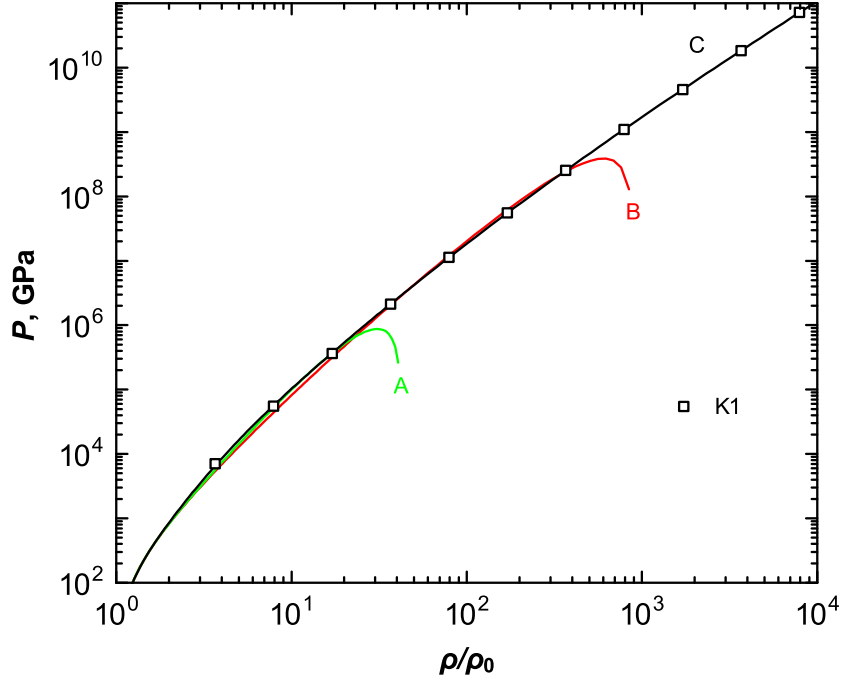}
\caption{The cold curve of tungsten at $T = 0$~K. Theoretical results: K1---the Thomas--Fermi model with corrections \cite{Kalitkin-Kuzmina-1975-eng}. Approximation curves: A---\!\!\cite{Kormer-Urlin-Popova-1961-eng}, B---\!\!\cite{Lomonosov-Thesis-2000}, C---this work.
\label{fig1}
}
\end{figure}

The cold-compression curve presented in figure~\ref{fig1} characterizes the accuracy of calculations of the pressure of tungsten at $T = 0$~K as a function of compression ratio, $P_c(\sigma)$. As one can see, the approximation relations proposed previously for $P_c(\sigma)$ are applicable in the region of compression ratios $\sigma \lesssim 20$ \cite{Kormer-Urlin-Popova-1961-eng} or $\sigma \lesssim 300$ \cite{Lomonosov-Thesis-2000} and lead to large errors or nonphysical results at higher densities. It should be noted that the applicability of relation~(\ref{Ecs1}) is restricted to the range in which the motion of electrons can be considered as non-relativistic~\cite{Kirzhnits-PhysUsp-1975-eng}: $5Am_uE_c/(3Z) \ll m_ec^2$, where $m_e$ is the electron mass and $c$ is the speed of light. This condition can be rewritten as $\sigma^{2/3} \ll 2Zm_ec^2/(5Am_uV_{0c}b_2$), which yields the following limit of applicability with respect to the compression ratio for the cold curve of tungsten obtained in this study: $\sigma \ll 3 \times 10^6$.

\begin{figure}[t]
\centering\includegraphics[width=0.75\columnwidth]{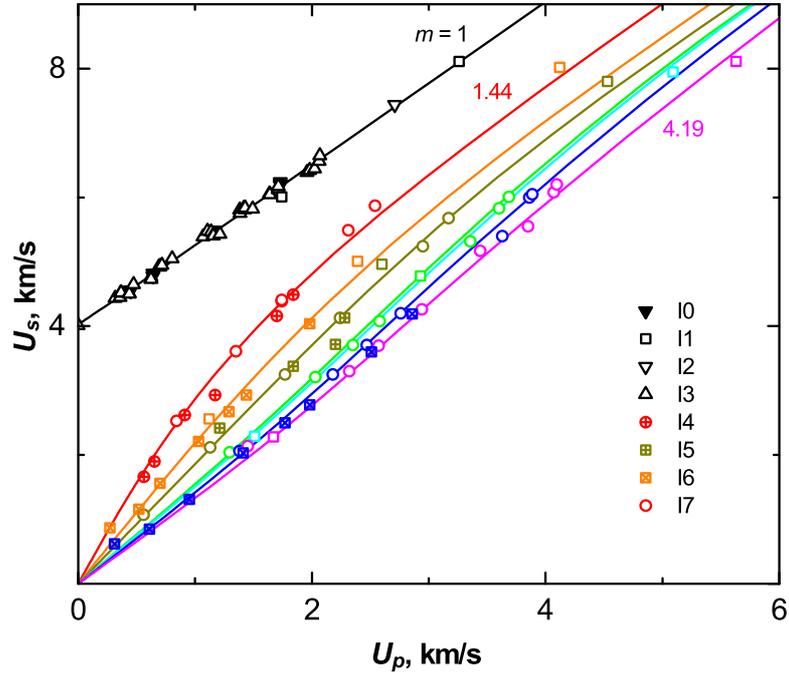}
\caption{The Hugoniots of tungsten samples with different initial porosity ($m$). Experimental data: 
I0---\!\!\cite{McQueen-Marsh-JAP-1960},
I1---\!\!\cite{Krupnikov-Brazhnik-Krupnikova-1962-eng},
I2---\!\!\cite{Jones-Isbell-Maiden-JAP-1966},
I3---\!\!\cite{LASL-1980},
I4---\!\!\cite{Boade-JAP-1969},
I5---\!\!\cite{Alekseev-Ratnikov-Ribakov-JAMTP-1971},
I6---\!\!\cite{Bakanova-Dudoladov-Sutulov-JAMTP-1974},
I7---\!\!\cite{Trunin-Simakov-Sutulov-Medvedev-Rogozkin-Fedorov-1989-eng}.
\label{fig2}
}
\end{figure}

\begin{figure}[t]
\centering\includegraphics[width=0.75\columnwidth]{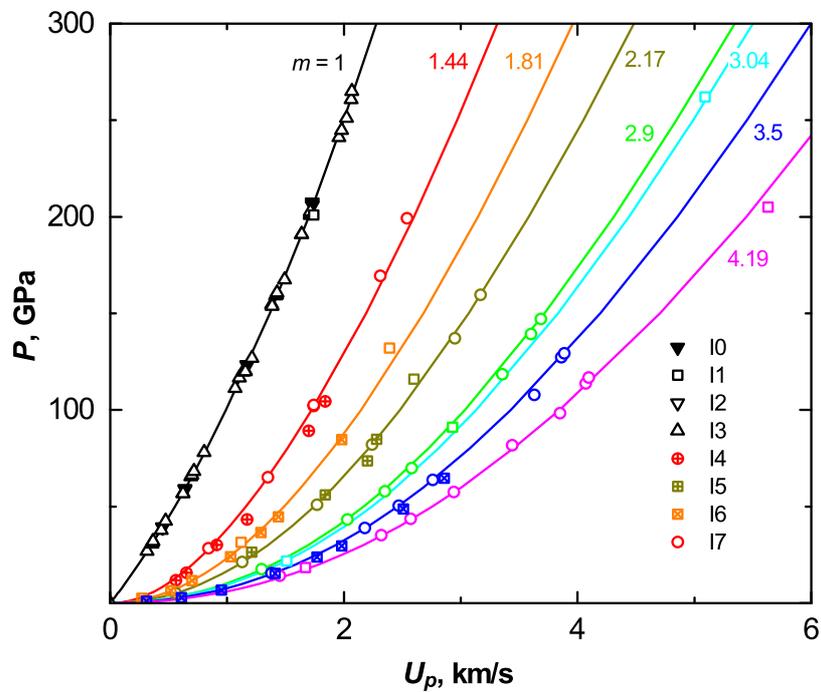}
\caption{The Hugoniots of tungsten samples with different initial porosity ($m$). Experimental data: markers are designated analogously to figure~\ref{fig2}.
\label{fig3}
}
\end{figure}

\begin{figure}[t]
\centering\includegraphics[width=0.75\columnwidth]{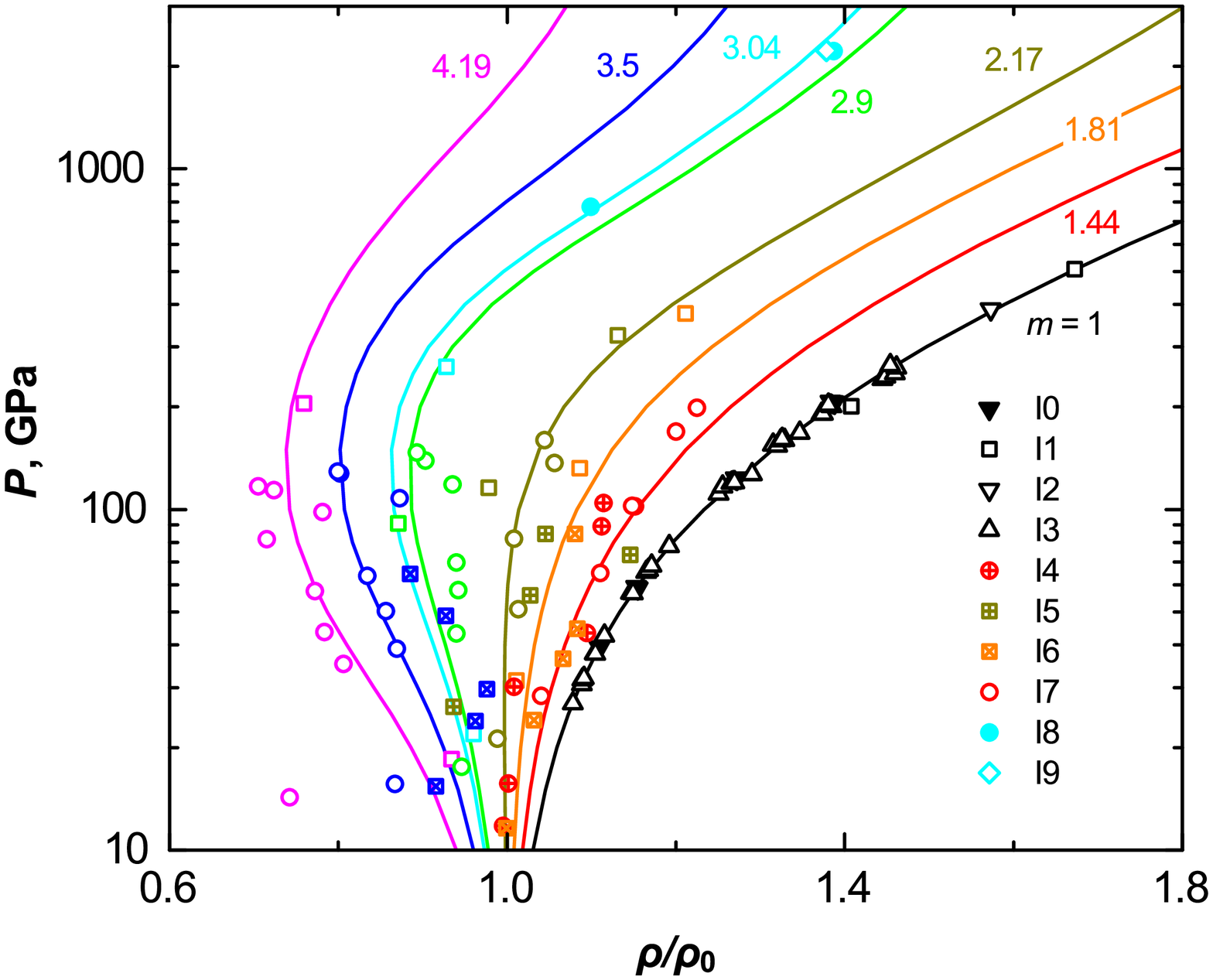}
\caption{The Hugoniots of tungsten samples with different initial porosity ($m$). Experimental data: markers are designated analogously to figure~\ref{fig2}, except for 
I8---\!\!\cite{Trunin-Medvedev-Funtikov-Podurets-Simakov-Sevastyanov-JETP-1989-eng},
I9---\!\!\cite{Trunin-1994-eng}.
\label{fig4}
}
\end{figure}

Shock compressibility of tungsten is studied in experiments with samples of different initial porosity $m=\rho_0/\rho_{00}$~\cite{McQueen-Marsh-JAP-1960, Krupnikov-Brazhnik-Krupnikova-1962-eng, Jones-Isbell-Maiden-JAP-1966, LASL-1980, Boade-JAP-1969, Alekseev-Ratnikov-Ribakov-JAMTP-1971, Bakanova-Dudoladov-Sutulov-JAMTP-1974, Trunin-Simakov-Sutulov-Medvedev-Rogozkin-Fedorov-1989-eng, Trunin-Medvedev-Funtikov-Podurets-Simakov-Sevastyanov-JETP-1989-eng, Trunin-1994-eng}, where $\rho_{00}$ is the initial density of samples, $\rho_0=V_0\!^{-1}$. A comparison of the results of calculations and the experimental data~\cite{McQueen-Marsh-JAP-1960, Krupnikov-Brazhnik-Krupnikova-1962-eng, Jones-Isbell-Maiden-JAP-1966, Boade-JAP-1969, Alekseev-Ratnikov-Ribakov-JAMTP-1971, Bakanova-Dudoladov-Sutulov-JAMTP-1974, LASL-1980, Trunin-Simakov-Sutulov-Medvedev-Rogozkin-Fedorov-1989-eng, Trunin-Medvedev-Funtikov-Podurets-Simakov-Sevastyanov-JETP-1989-eng, Trunin-1994-eng} presented in figures~\ref{fig2}--\ref{fig4} shows that the proposed EOS provides for a reliable description of the metal properties in wide ranges of shock and particle velocities ($U_s$ and $U_p$ respectively), pressures and compression ratios.

Thus the EOS proposed for tungsten may be used effectively in numerical simulations of processes of interaction of intense energy fluxes with the material.

\ack
The work is supported by the Russian Science Foundation grant No.\,14-50-00124.

\section*{References}

\begin{thebibliography}{10}
\expandafter\ifx\csname url\endcsname\relax
  \def\url#1{{\tt #1}}\fi
\expandafter\ifx\csname urlprefix\endcsname\relax\def\urlprefix{URL }\fi
\providecommand{\eprint}[2][]{\url{#2}}

\bibitem{Bushman-Fortov-Kanel-Ni-1993}
Bushman A~V, Fortov V~E, Kanel' G~I and Ni A~L 1993 {\em Intense Dynamic
  Loading of Condensed Matter\/} (Washington: Taylor \& Francis)

\bibitem{Russkikh-Baksht-Labetskii-Levashov-Tkachenko-Khishchenko-Shishlov-Fedyunin-Chaikovskii-PPR-2006-eng}
Russkikh A~G, Baksht R~B, Labetskii A~Y, Levashov P~R, Tkachenko S~I,
  Khishchenko K~V, Shishlov A~V, Fedyunin A~V and Chaikovskii S~A 2006 {\em
  Plasma Phys. Rep.\/} {\bf 32} 823--835

\bibitem{Rousskikh-Baksht-Chaikovsky-Fedunin-Khishchenko-Labetsky-Levashov-Shishlov-Tkachenko-2006}
Rousskikh A~G, Baksht R~B, Chaikovsky S~A, Fedunin A~V, Khishchenko K~V,
  Labetsky A~Y, Levashov P~R, Shishlov A~V and Tkachenko S~I 2006 {\em IEEE
  Trans. Plasma Sci.\/} {\bf 34} 2232--2238

\bibitem{Senchenko-Belikov-Popov-2015}
Senchenko V~N, Belikov R~S and Popov V~S 2015 {\em J. Phys.: Conf. Series\/}
  {\bf 653} 012100

\bibitem{Rakhel-Korobenko-Savvatimski-Fortov-IJT-2004}
Rakhel A~D, Korobenko V~N, Savvatimski A~I and Fortov V~E 2004 {\em Int. J.
  Thermophys.\/} {\bf 25} 1203--1214

\bibitem{Oreshkin-Baksht-Labetsky-Rousskikh-Shishlov-Levashov-Khishchenko-Glazyrin-TP-2004-eng}
Oreshkin V~I, Baksht R~B, Labetsky A~Y, Rousskikh A~G, Shishlov A~V, Levashov
  P~R, Khishchenko K~V and Glazyrin I~V 2004 {\em Tech. Phys.\/} {\bf 49}
  843--848

\bibitem{Tkachenko-Khishchenko-Levashov-IJT-2005}
Tkachenko S~I, Khishchenko K~V and Levashov P~R 2005 {\em Int. J.
  Thermophys.\/} {\bf 26} 1167--1179

\bibitem{Povarnitsyn-Khishchenko-Levashov-IJImpEng-2006}
Povarnitsyn M~E, Khishchenko K~V and Levashov P~R 2006 {\em Int. J. Impact
  Eng.\/} {\bf 33} 625--633

\bibitem{Tkachenko-Levashov-Khishchenko-JPA-2006}
Tkachenko S~I, Levashov P~R and Khishchenko K~V 2006 {\em J. Phys. A: Math.
  Gen.\/} {\bf 39} 7597--7603

\bibitem{Oreshkin-Khishchenko-Levashov-Rousskikh-Chaikovskii-HT-2012}
Oreshkin V~I, Khishchenko K~V, Levashov P~R, Rousskikh A~G and Chaikovskii S~A
  2012 {\em High Temp.\/} {\bf 50} 584--595

\bibitem{Andreev-Povarnitsyn-Veysman-Faenov-Levashov-Khishchenko-Pikuz-Magunov-Rosmej-Blazevic-Pelka-Schaumann-Schollmeier-Roth-LPB-2015}
Andreev N~E, Povarnitsyn M~E, Veysman M~E, Faenov A~Y, Levashov P~R,
  Khishchenko K~V, Pikuz T~A, Magunov A~I, Rosmej O~N, Blazevic A, Pelka A,
  Schaumann G, Schollmeier M and Roth M 2015 {\em Laser Part. Beams\/} {\bf 33}
  541--550

\bibitem{Kormer-Urlin-Popova-1961-eng}
Kormer S~B, Urlin V~D and Popova L~T 1961 {\em Fiz. Tverd. Tela\/} {\bf 3}
  2131--2140

\bibitem{Bushman-Fortov-Lomonosov-1991-proc}
Bushman A~V, Fortov V~E and Lomonosov I~V 1991 {\em High Pressure Equations of
  State: Theory \& Applications\/} ed Eliezer S and Ricci R~A (Amsterdam:
  North-Holland) pp 249--262

\bibitem{Bushman-Lomonosov-Fortov-1992-metals-eng}
Bushman A~V, Lomonosov I~V and Fortov V~E 1992 {\em Equations of State of
  Metals at High Energy Densities\/} (Chernogolovka: ICP RAS)

\bibitem{Lomonosov-Bushman-Fortov-Khishenko-1994}
Lomonosov I~V, Bushman A~V, Fortov V~E and Khishenko K~V 1994 {\em
  High-Pressure Science and Technology---1993\/} ed Schmidt S~C, Shaner J~W,
  Samara G~A and Ross M (New York: AIP Press) pp 133--136

\bibitem{Lomonosov-Thesis-2000}
Lomonosov I~V 2000 {D}.Sc. thesis Institute of Problems of Chemical Physics
  RAS, Chernogolovka

\bibitem{Tkachenko-Khishchenko-Vorobev-Levashov-Lomonosov-Fortov-HT-2001-eng}
Tkachenko S~I, Khishchenko K~V, Vorob'ev V~S, Levashov P~R, Lomonosov I~V and
  Fortov V~E 2001 {\em High Temp.\/} {\bf 39} 674--687

\bibitem{Lomonosov-Fortov-Khishchenko-Levashov-2002-article}
Lomonosov I~V, Fortov V~E, Khishchenko K~V and Levashov P~R 2002 {\em AIP Conf.
  Proc.\/} {\bf 620} 111--114

\bibitem{Fortov-Lomonosov-PhysUsp-2014}
Fortov V~E and Lomonosov I~V 2014 {\em Phys. Usp.\/} {\bf 57} 219--233

\bibitem{Khishchenko-TPL-2004-eng}
Khishchenko K~V 2004 {\em Tech. Phys. Lett.\/} {\bf 30} 829--831

\bibitem{Kalitkin-Govorukhina-1965}
Kalitkin N~N and Govorukhina I~A 1965 {\em Sov. Phys.--Solid State\/} {\bf 7}
  287

\bibitem{Landau-Lifshitz-V-1980-eng}
Landau L~D and Lifshitz E~M 1980 {\em Course of Theoretical Physics. Vol.~5.
  Statistical Physics. Part~1\/} (Oxford: Pergamon)

\bibitem{Kalitkin-Kuzmina-1975-eng}
Kalitkin N~N and Kuzmina L~V 1975 {\em Tables of Thermodynamic Functions of
  Matter at High Concentration of Energy. Preprint No.~35\/} (Moscow: Institute
  of Applied Mathematics of the Academy of Sciences USSR)

\bibitem{Bushman-Lomonosov-Fortov-Khishchenko-Zhernokletov-Sutulov-JETP-1996}
Bushman A~V, Lomonosov I~V, Fortov V~E, Khishchenko K~V, Zhernokletov M~V and
  Sutulov Y~N 1996 {\em JETP\/} {\bf 82} 895--899

\bibitem{Kirzhnits-PhysUsp-1975-eng}
Kirzhnits D~A 1971 {\em Sov. Phys. Usp.\/} {\bf 14} 512

\bibitem{McQueen-Marsh-JAP-1960}
McQueen R~G and Marsh S~P 1960 {\em J. Appl. Phys.\/} {\bf 31} 1253--1269

\bibitem{Krupnikov-Brazhnik-Krupnikova-1962-eng}
Krupnikov K~K, Brazhnik M~I and Krupnikova V~P 1962 {\em Zh. Eksp. Teor.
  Fiz.\/} {\bf 42} 675--685

\bibitem{Jones-Isbell-Maiden-JAP-1966}
Jones A~H, Isbell W~H and Maiden C~J 1966 {\em J. Appl. Phys.\/} {\bf 37}
  3493--3499

\bibitem{LASL-1980}
Marsh S~P (ed) 1980 {\em LASL Shock Hugoniot Data\/} (Berkeley, CA: University
  of California Press)

\bibitem{Boade-JAP-1969}
Boade R~R 1969 {\em J. Appl. Phys.\/} {\bf 40} 3781--3792

\bibitem{Alekseev-Ratnikov-Ribakov-JAMTP-1971}
Alekseev Y~L, Ratnikov B~P and Ribakov A~P 1971 {\em J. Appl. Mech. Tech.
  Phys.\/} {\bf 12} 257--263

\bibitem{Bakanova-Dudoladov-Sutulov-JAMTP-1974}
Bakanova A~A, Dudoladov I~P and Sutulov Y~N 1974 {\em J. Appl. Mech. Tech.
  Phys.\/} {\bf 15} 241--245

\bibitem{Trunin-Simakov-Sutulov-Medvedev-Rogozkin-Fedorov-1989-eng}
Trunin R~F, Simakov G~V, Sutulov Y~N, Medvedev A~B, Rogozkin B~D and Fedorov
  Y~E 1989 {\em Zh. Eksp. Teor. Fiz.\/} {\bf 96} 1024--1038

\bibitem{Trunin-Medvedev-Funtikov-Podurets-Simakov-Sevastyanov-JETP-1989-eng}
Trunin R~F, Medvedev A~B, Funtikov A~I, Podurets M~A, Simakov G~V and
  Sevast'yanov A~G 1989 {\em Sov. Phys. JETP\/} {\bf 68} 356--361

\bibitem{Trunin-1994-eng}
Trunin R~F 1994 {\em Phys. Usp.\/} {\bf 37} 1123--1146

\end{thebibliography}
\providecommand{\newblock}{}

\end{document}